\documentclass[runningheads]{llncs}
\usepackage{graphicx}
\begin{document}
\title{The current design and implementation of the AstroDS Data Aggregation Service\thanks{Supported by the Russian Science Foundation, grant \#18-41-06003.}}
\author{Minh-Duc Nguyen\inst{1}\orcidID{0000-0002-5003-3623} \and
Alexander Kryukov\inst{1}\orcidID{0000-0002-1624-6131} \and
Andrey Mikhailov\inst{2}\orcidID{0000-0003-4057-4511}}
\authorrunning{M.-D. Nguyen et al.}
\institute{
Skobeltsyn Institute of Nuclear Physics, Lomonosov Moscow State University, Moscow, Russia \email{nguyendmitri@gmail.com}
\and
Matrosov Institute for System Dynamics and Control Theory of Siberian Branch of Russian Academy of Sciences Irkutsk, Russia
}
\maketitle              % typeset the header of the contribution
\begin{abstract}
AstroDS is a distributed storage for Cosmic Ray Astrophysics. The primary goal of Astro DS is to gather data measured by the instruments of various physical experiments such as TAIGA, TUNKA, KASCADE into global storage and provide the users with a standardized user-friendly interface to search for the datasets that match certain conditions. AstroDS consists of a set of distributed microservices components that communicate with each other through the Internet via REST API. The core component of AstroDS is the Data Aggregation Service that orchestrates other components to provide access to data. The development process of AstroDS started in 2019. This paper describes the current design and implementation of the Data Aggregation Service and also the benefits it brings to the astrophysical community in the early state.

\end{abstract}
\section{Introduction}

Russian–German Astroparticle Data Life Cycle Initiative~\cite{astrods_initiative} is an international project whose aim is to develop an open science system called ASTROPARTICLE.ONLINE that enables users to publish, store, search, select, and analyze astroparticle data that are coming from various experiments located worldwide. One of the main parts of the system is the Astroparticle Physics Distributed Data Storage~\cite{astrods_2019} (AstroDS) that gathers data measured by the instruments of physical experiments such as TAIGA~\cite{taiga}, KASCADE-Grande~\cite{kascade} into global storage and allows users to search for a specific dataset and retrieve it via a standardized storage-independent API.

AstroDS consists of a set of distributed microservices that interacts with each other to provide a smooth experience for users in data acquisition. One of the core services of AstroDS is the Data Aggregation Service (the Service) that coordinates others to fulfill data queries from users~\cite{aggregator_2019}. This paper describes the current design and implementation of the Service.

\section{Design}

\begin{figure}
\includegraphics[width=\textwidth]{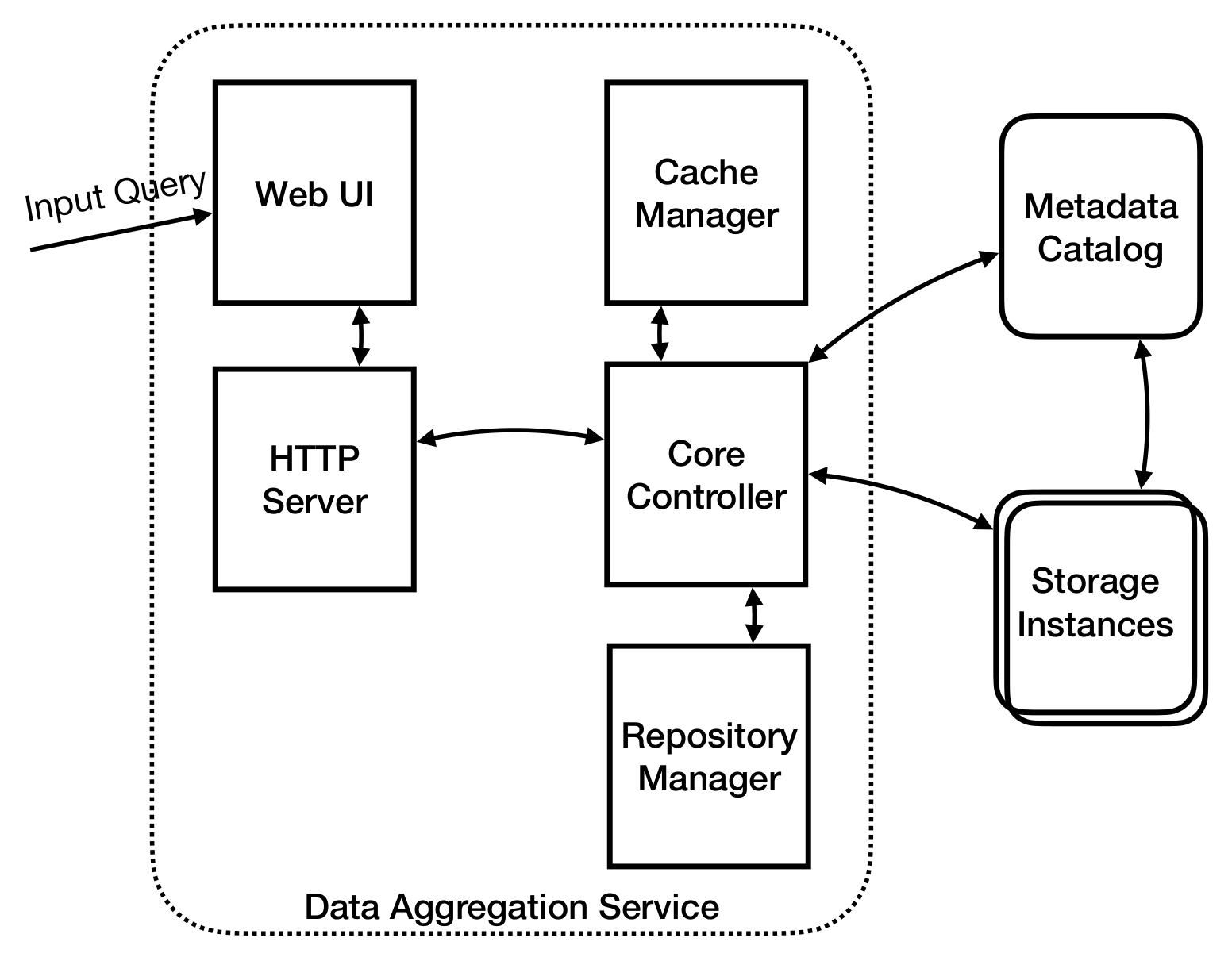}
\caption{The architecture design of the Data Aggregation Service} \label{figure_aggregator_design}
\end{figure}

The overall view of the Service's components is presented in Fig.~\ref{figure_aggregator_design}. On the frontend side, through the Web User Interface, the Service receives data lookup queries from users. The content of a query contains a set of search filters defined by a user. For example, the query can contain IACT01, which is the code name of a facility of the TAIGA experiment, two timestamps identifying the start and the end of a certain period, the weather condition when the data were collected at the experiment's site.

On the backend side, the Web User Interface is served by an HTTP Server. The HTTP Server serves as a proxy layer that passes the queries to different instances of the Service. Within each instance of the Service, the Core Controller is responsible for processing queries. Upon receiving a query, the Core Controller always checks with the Cache Manager if there is a response already generated for the query. To uniquely identify a query, the Core Controller uses the MD5 algorithm to calculate the hash of the query using its content. A response to a query is a list of files containing data that matches the filters defined in the query. If the response to a query is not cached, the Core Controller will forward the query to the Metadata Catalog that stores the metadata of all available datasets. The Metadata Catalog, in turn, makes a search query against its database and returns a response to the Core Controller. After receiving the response from the Metadata Catalog, the Core Controller passes it to the Cache Manager for caching and requests the Repository Manager to generate the files in the response using data from the Storage Instances each of which is mounted to the machine where the Service is running. File generation is carried out asynchronously.

\section{Implementation}

The Web UI was implemented as a single-page application based on React~\cite{react} and Material UI~\cite{material_ui} packed into a Docker container. In the Web UI, the user can create a data query using the search menu. The GraphQL query language~\cite{graphql} was used to implement the query. The structure of the query reflects the structure of entities that define the data of a physical experiment such as experiment site, facility name, instrument name, detector, data channel, etc. To compose a query, the user can define a list of filters and corresponding values. Each facility of a physical experiment has its own set of filters. When the user chooses multiple facilities, only the common filters available for all facilities are shown, facility-specific filters are hidden.

\begin{figure}
\includegraphics[width=\textwidth]{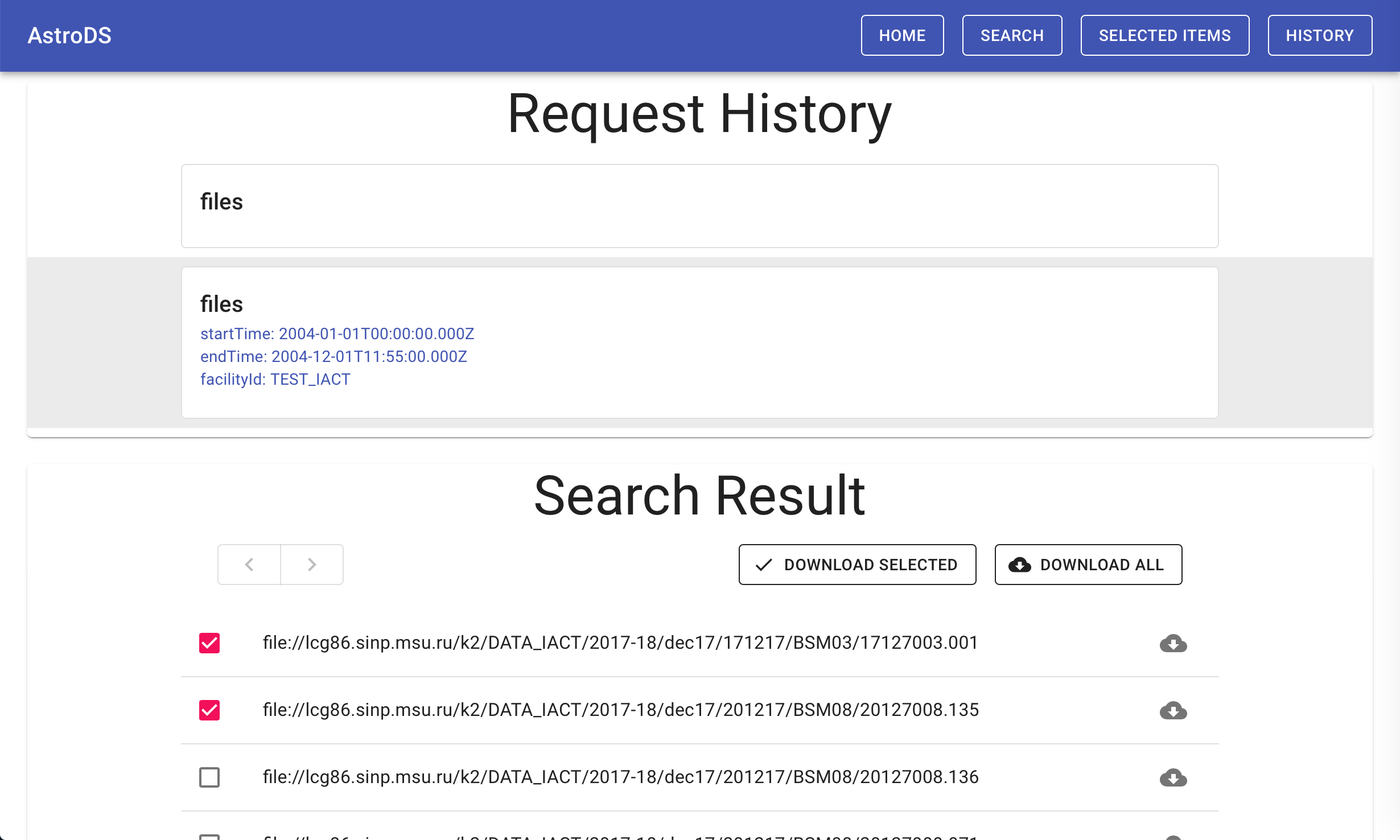}
\caption{The architecture design of the Data Aggregation Service} \label{figure_web_ui}
\end{figure}

Since the history of all queries made by the user is stored in the local storage of the browser, it is possible to review the response to a query. While reviewing the responses, the user can make a list of files from different queries just by selecting them. The selected files later can be downloaded as a single archive. An illustration of the Web UI is shown in Fig.~\ref{figure_web_ui}

The Core Controller was implemented as a microservice based on the Django framework~\cite{django} and packed into a Docker container. The whole communication between the Web UI and the Core Controller is done through a set of REST API~\cite{rest_api} endpoints, each of which is responsible for a type of queries. Data lookup queries are formed using the GraphQL syntax and sent to the Core Controller as POST-requests; others are standard REST API requests. One exception is the file generation queries. Since files are generated asynchronously, when the files for a query are generated, the Core Controller sends a short text message to the Web UI via the WebSocket channel~\cite{websocket}. The Web UI can also check the current status of a file generation query using a GET-request on the initialization stage.

All queries are cached by the Cache Manager to reduce the response time. The cache was implemented as a big hash table where the key is the MD5 hash of a query, and the value contains the response to the query and other metadata such as creation time, last used timestamp. Queries are discarded from the cache by the Least Frequently Used (LFU) policy: the Cache Manager counts how often a query is used; those that are used least often are discarded first. Currently, the expiration time of a query is seven days. 

The Repository Manager was implemented as a separate microservice with its REST API. After receiving a file generation request from the Core Controller, the Repository Manager creates an empty CernVM-FS repository, copies the original files from the storage instances to the newly created repository, and publishes it. When a repository is ready, the Repository Manager sends a short text message to the Core Controller via the WebSocket channel.

To implement the communication via WebSocket, the RabbitMQ Message Broker~\cite{rabbitmq} was used. Text messages are formatted using the STOMP protocol~\cite{stomp}.

\section{Current Status}

Facilities of the physical experiments mentioned above (TAIGA, KASCADE-Grande) were successfully integrated into the AstroDS system. Using the Web UI of the Data Aggregation Service, users can already make a complex query that combines data from different facilities of the experiments. The KASCADE-Grande experiment has its own service similar to the Repository Manager, so components of the Data Aggregation Service work with that service to generate files. There are ongoing works to make the Data Aggregation Service stable in production use. In the future, there are plans to integrate more physical experiments into the AstroDS platforms and implement more filters that give users a flexible way to search for data. There are also plans to conduct outreach activities to increase the number of users.

\end{document}